\documentclass[letterpaper,10pt]{extarticle}  
\usepackage{import}
\usepackage{local}
\usepackage[left=.65in,top=.65in,bottom=.65in,right=.65in]{geometry}
\usepackage{lipsum} 

\usepackage[version=3]{mhchem}
\usepackage{amsmath}
\usepackage{epstopdf}
\usepackage{hyperref}
\usepackage{array}
\usepackage{balance}
\usepackage{times,mathptmx}
\usepackage{graphicx} 
\usepackage{lastpage}

\newcommand{\bra}[1]{\langle #1 |}                 
\newcommand{\ket}[1]{| #1 \rangle}                 

\title{Bending relaxation of H$_{2}$O by collision with
	\textit{para}- and \textit{ortho}-\ce{H2}}

\author{%
\textbf{Ricardo Manuel García-Vázquez \orcidlink{0000-0002-3786-8028},\textcolor{Accent}{\textsuperscript{1}} %
Alexandre Faure \orcidlink{0000-0001-7199-2535},\textcolor{Accent}{\textsuperscript{2}} %
Thierry Stoecklin \orcidlink{0000-0002-1349-8106}\textcolor{Accent}{\textsuperscript{1,*}} }\\
\begin{small}\textcolor{Accent}{\textsuperscript{1}}UMR5255-CNRS, Universit\'e de Bordeaux, 351 cours de la lib\'eration , F-33405 Talence, France. \\ 
\textcolor{Accent}{\textsuperscript{2}}Univ. Grenoble Alpes, CNRS, IPAG,
F-38000 Grenoble, France. \\
\textcolor{Accent}{\textsuperscript{*}}Correspondence: \textcolor{Accent}{thierry.stoecklin@u-bordeaux.fr} \\ \end{small}
}

\date{14/11/2023}
\begin{document}
\maketitle
\thispagestyle{empty}

\section{Abstract}

\begin{doublespacing}

\noindent
\textbf{\textcolor{Accent}{We extend our recent theoretical work on the
		bending relaxation of \ce{H2O} in collisions with \ce{H2} by
		including the three water modes of vibration coupled with rotation,
		as well as the rotation of \ce{H2}. Our full quantum close-coupling
		method (excluding the H$ _{2} $ vibration) is combined with a high-accuracy nine-dimensional potential
		energy surface. The collisions of \textit{para}-\ce{H2O} and
		\textit{ortho}-\ce{H2O} with the two spin modifications of \ce{H2}
		are considered and compared for several initial states of
		\ce{H2O}. The convergence of the results as a function of the size
		of the rotational basis set of the two colliders is discussed. In
		particular, near-resonant energy transfer between \ce{H2O} and
		\ce{H2} is found to control the vibrational relaxation process, with
		a dominant contribution of transitions with $\Delta j_2=j_{2}^{i}-j_{2}^{f}=+2, +4$, $ j_{2}^{i} $ and $ j_{2}^{f} $ being respectively the \ce{H2} initial and final rotational quantum numbers. Finally, the calculated value of the \ce{H2O} bending
		relaxation rate coefficient at 295~K is found to be in excellent agreement with its experimental estimate.}}

\section{Introduction}
\label{introduction}

Until very recently, the modeling of molecular excitation in the
interstellar medium only took into account the rotation of the
molecules detected, due to the cold temperatures ($<$ 100~K)
prevailing in interstellar clouds. However, some of the most abundant
triatomic interstellar molecules, such as \ce{HCN}
\cite{HCN-cernicharo:11} and \ce{C3} \cite{C3-cernicharo:00}, have
been detected in highly excited vibrational states in warmer
environments, e.g. in star-forming regions and in the envelopes of
giant stars. The large quantity of water present in the Earth's
atmosphere made it very difficult to detect \ce{H2O} in
ro-vibrationally excited states but the construction of the Atacama
Large Millimeter/submillimeter Array (ALMA) interferometer partially
overcame these constraints and enabled the first measurements of water
transitions into vibrationally excited states, up to $(\nu_{1}, \nu_{2}, \nu_{3})=(0, 1, 1)$, as
part of the Atomium project \cite{baudry2023atomium}. More recently,
the launch of the James Webb Space Telescope (JWST) has not only freed
us from the Earth's atmosphere, but has also widened the window of
detection to vibrational wavelengths, with e.g. the very recent
detection of water vapor in the terrestrial planet-forming region of a
protoplanetary disk \cite{Perotti2023}.

Collisional rate coefficients of \ce{H2} with water in vibrationally
excited levels are then urgently needed by the astrochemists. On the
theoretical side, the first quantum calculations dedicated to
collisions involving vibrationally excited polyatomic molecules were
performed by Dagdigian and Alexander in 2013
\cite{CH2-He-Dagdigian-Alexander-2013} for the umbrella modes of
\ce{CH3} colliding with He and by Loreau and Van der Avoird in 2015
for the umbrella modes of \ce{NH3} in collisions with
He\cite{NH3-Van-der-Avoird-2015} . We developed our own approach to
treat the bending relaxation of a linear triatomic molecule colliding
with an atom shortly after and applied it to the
He-\ce{HCN}\cite{stoecklin:13,Denis:13}, He-\ce{C3}
\cite{Mogren:14,Denis:14,stoecklin-2015-he-c3}and He-\ce{DCN}
\cite{Denis:15} collisions. This method uses the internal coordinates
Hamiltonian developed by Handy and Tennyson \cite{DVR3D-TENNYSON-2004}
restricted to its rigid bender form to compute the triatomic molecule
ro-vibrational bound state energies and wave functions.  We extended
this approach recently to collisions involving an atom and a non
linear symmetric triatomic molecule and applied it to the bending
relaxation of H$_{2}$O colliding with \textit{ para}-H$_{2}$
\cite{H2-H2O-Stoecklin-2019}, \textit{ para}-H$_{2}$ being treated as
fictitious atom. In 2021 the team of Guo used a very similar approach
to treat the vibrational relaxation of water, including the bending
and stretching modes, in collisions with Cl \cite{H2O-Cl-Guo-2021} and
Ar \cite{Ar-H2O-2023-Guo}. They used Radau coordinates instead of Bond
length Bond angle coordinates to obtain the triatomic energies and
wave functions.

For many years the only calculations available regarding the \ce{H2 +
	H2O } collision and taking into account the rotation of \ce{H2 } as
well as the bending and rotation of \ce{H2O} were the Quasi-Classical
Trajectory (QCT) calculations performed by Faure \textit{et~al.}
\cite{H2O-vib-faure-2005} using the full-dimensional (9D) potential
energy surface (PES), fully described in Valiron \textit{et~al.}
\cite{Valiron-2008}. The first quantum calculations of the bending
relaxation of water by collisions with \ce{H2 } taking into account
the rotation of \ce{H2} were performed by Wiesenfeld in 2021
\cite{H2-H2O-2021-Wiesenfeld}. These calculations however did not
include yet the coupling between the bending and rotation of
\ce{H2O}. The object of the present study is then to extend our
previous work on this system \cite{H2-H2O-Stoecklin-2019} by taking
also into account both the rotation of \ce{H2} and the two stretching
modes of \ce{H2O}. To this aim we use the same 9D PES developped by
Valiron \textit{et~al.} \cite{Valiron-2008}. The paper is organised as
follows: In section \nameref{Meth} the main elements of the method are
briefly reminded and the modifications necessary to perform the
dynamics calculations are discussed. In section \nameref{CalRes} the
results are presented and the bending relaxation rate coefficient is
compared to the experimental data available. In section
\nameref{Concl} a few conclusions are given.

\section{Method }\label{Meth}

The rigid rotor Close Coupling equations for the collisions of \ce{H2
	+ H2O} were given long ago in detail by Phillips \textit{et~al.}
\cite{H2-H2O-General-CS-Green-1995} and we refer the interested reader
to this paper. The extension of the theory to vibrating molecules is
straightforward and only a brief account of the main steps of the
calculations will be given in section \nameref{cc}. The Schrödinger
equation of the system is expressed in the $(X',Y',Z')$ Space fixed frame which is
obtained by a simple rotation of the inertial \ce{H2O} $(X,Y,Z)$ frame
illustrated in Fig.~1 where $ \overrightarrow{R} $ is the
intermolecular vector between the centers of mass of \ce{H2O} and
\ce{H2} and $ (R,\theta_{R},\varphi_{R}) $ are its spherical
coordinates in this frame where the relative interaction potential
briefly described in the following section is also computed. As can be
seen on this figure the water molecule is lying in the XOZ molecular
plane while the $ C_{2} $ Z axis is bisecting the bending angle.

\subsection{Potential energy surface}\label{sec:pes} 
We use an 8D restriction to non vibrating \ce{H2} of the 9D PES of Valiron \textit{et~al.} \cite{Valiron-2008} which
is expanded in the rigid rotor Green's angular basis set
\cite{H2-H2Orot-Green-1996}
\begin{equation}
	\begin{split}
		t_{p_{1} p_{2} p} ^{q_{1}}(\theta_{2},\varphi_{2},\theta_{R},\varphi_{R}) = 
		\dfrac{\alpha^{q_1}_{p_1p_2p}}{(1+\delta_{q_{1}r_{1}})}\sum_{r_{1},r_{2}}
		\left (\begin{array}{ccc}
			p_{1 }& p_{2} & p \\ 
			r_{1} & r_{2} & r 
		\end{array}\right )\times
		Y_{p_{2}}^{r_{2}}(\theta_{2},\varphi_{2}) Y_{p}^{r}(\theta_{R},\varphi_{R})
		[\delta_{q_{1}r_{1}}+(-1)^{(p_{1}+q_{1}+p_{2}+p)}\delta_{-q_{1}r_{1}}]
		\label{eqn:Green-basis}
	\end{split} 
\end{equation}
for each value of the \ce{H2O} Radau coordinates $ (R_{1},R_{2},\alpha
)$ and the intermolecular distance R:
\begin{equation}
	\begin{split}
		V(R_{1},R_{2},\alpha,R,\theta_{2},\varphi_{2},\theta_{R},\varphi_{R}) =
			\sum_{p_{1} q_{1}p_{2} p}	V_{p_{1} p_{2} p}^{q_{1}}(R_{1},R_{2},\alpha,R)
		\, t_{p_{1} p_{2} p} ^{q_{1}}(\theta_{2},\varphi_{2},\theta_{R},\varphi_{R}).
	\end{split} 
\end{equation}\label{eqn:expension}
In Eq.~(1), the normalization factor is \cite{Valiron-2008}:
\begin{equation}
	\alpha^{q_1}_{p_1p_2p}=[2(1+\delta_{q_{1}0})^{-1}(2p_1+1)^{-1}]^{-1/2}.
\end{equation}

\begin{figure}
	\centering
	\includegraphics[width=0.4\textwidth]{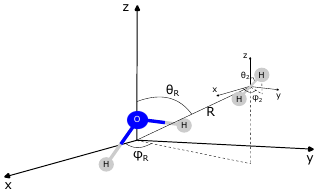}
	\caption{Coordinates used for the H$ _{2} $-H$ _{2} $O
		collision. The water molecule lies in the XOZ plane where O the \ce{H2O} center of mass and the
		Z axis is bisecting the bending angle. \label{coord}}
\end{figure}

\begin{figure}
	\centering
	\includegraphics[width=0.3\textwidth]{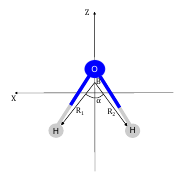}
	\caption{Coordinates used for the calculation of the H$ _{2}$O bound states. B is the Radau canonical point of \cf{H2O}, $R_1$ and $R_2$ are the modulus of the vectors that defines the position of the two hydrogen atoms with respect to the B point and $\alpha$ is the angle between these two vectors.}  \label{coord2}
\end{figure}

\subsection{Calculations of the ro-vibrational bound states of \ce{H2O}  }\label{h2o-bs}

Many papers have been devoted to the study of Energy levels and spectra of the water molecule from the experimental and theoretical point of view. Very accurate results can be found, for example, in Ref.~\citenum{Polyansky2013,Polyansky2018}, and we encourage the people interested in the use of an extensive list of energy levels to refer to these papers. The first objective of this section is the determination of the \ce{H2O} rovibrational wave functions which will be used for performing the dynamics calculations. The second objective is  to point out the main differences between our new approach and the one used in our previous work on the system\cite{H2-H2O-Stoecklin-2019} as well as with the widely used DVR3D method \cite{DVR3D-TENNYSON-2004}.

In our previous work dedicated to the rigid bender inelastic collision
of H$_{2}$O colliding with \textit{ para}-H$_{2}$
\cite{H2-H2O-Stoecklin-2019} we used bond length bond angle
coordinates together with the Hamiltonian of Tennyson and Sutcliffe
\cite{Tennyson-IJQC-92} to describe the internal motion of the water
molecule. The rigid bender Hamiltonian equations were solved in the X
bisector embedding frame and the solutions rotated in the X-Z plane by
an angle of $\dfrac{\pi}{2}$ in order to perform the dynamics .

In the present work, as we need to describe equally the stretching,
bending and rotation of \ce{H2O}, we use instead Radau coordinates
($R_1$, $R_2$, $\alpha$) (see Fig. \ref{coord2}) and the Z bisector embedding. This choice
simplifies the hamiltonian by cancelling the radial cross derivatives
that appears in the bond length bond angle Hamiltonian and also
eliminating the need of the rotation step after diagonalization. It
furthermore offers an excellent compromise of accuracy and computer
time to calculate the \ce{H2O} ro-vibrational bound states energies
and wave functions as demonstrated by Tennyson
\cite{DVR3D-TENNYSON-2004}. In what follows we give the details of our
implementation of the method for solving these equations as it differs
in several respects from the approach used in the code DVR3D
\cite{DVR3D-TENNYSON-2004} like for example using Potential Optimised
Discrete Variable Representation (PODVR)
\cite{H2O-Cl-Guo-2021,Optimized-DVR-Clary-1992}.

The \ce{H2O } Hamiltonian takes the following form in Radau coordinates $ (R_{1},R_{2},\alpha) $ and atomic units\cite{Wang2017}:
\begin{equation}\label{eq:ABC_hamil}
	\hat{H}_{ABC} = - \dfrac{1}{2\mu_1}\dfrac{\partial^2}{\partial
		R_1^2} - \dfrac{1}{2\mu_2}\dfrac{\partial^2}{\partial R_2^2}
	+ \hat{T}_{\alpha} + \hat{T}_{vr} + V_{ABC}(R_1,R_2,\alpha)
\end{equation} 
where $\hat{T}_{\alpha}$ and $\hat{T}_{vr}$ are respectively the
kinetic operators associated with the Radau angle $ \alpha$ and with
the coupling between vibration and rotation as defined in
Ref.~[\citenum{Wang2017}] while the $\mu_i$ are the reduced masses
associated with the $R_i$ coordinates \cite{DVR3D-TENNYSON-2004}. The
expression (\ref{eq:ABC_hamil} ) is as usual re-written in terms of
single radial coordinate hamiltonian of $R_1$ and $R_2$ as:
\begin{equation}\label{hamil}
	\hat{H}_{ABC} = \hat{h}_1(R_1) +\hat{h}_2(R_2) + \hat{	T }_{\alpha} +  \hat{T}_{vr} + \hat{V}_{res}(R_1,R_2,\alpha)
\end{equation} 
where $\hat{h}_i, i=1,2$ are defined as:
\begin{equation}
	\hat{h}_i(R_i) = -  \dfrac{1}{2\mu_i}\dfrac{\partial^2}{\partial R_i^2} + \hat{V}_i(R_i)
\end{equation}
These reference hamiltonians are similar to those used in
Ref.~[\citenum{H2O-Cl-Guo-2021}] with the difference that we use the
reduced masses $\mu_1$ and $\mu_2$ instead of the atomic masses $m_A$
and $m_C$. The $\hat{V}_i(R_i)$ and $\hat{V}_{res}(R_1,R_2,\alpha)$
potentials are defined in the same way as in Ref.~[\citenum{H2O-Cl-Guo-2021}]:
\begin{eqnarray}
	\hat{V}_1(R_1) = \hat{V}_{ABC}(R_1,R_{2}^{e},\alpha^{e})\\
	\hat{V}_2(R_2) = \hat{V}_{ABC}(R_{1}^{e},R_2,\alpha^{e})
\end{eqnarray}  
\begin{equation}
	\hat{V}_{res}(R_1,R_2,\alpha)  = \hat{V}_{ABC}(R_1,R_2,\alpha) - \hat{V}_1(R_1) - \hat{V}_2(R_2)
\end{equation}
where $(R_{1}^{e},R_{2}^{e},\alpha^{e})$ are the values of the Radau coordinates for the equilibrium geometry of the ABC molecule. 

For the sake of simplicity the  \ce{H2O} rotational angular momentum $ j_{1} $ as well as its space fixed Z axis projection $ m_{j_{1}} $  will be denoted only $ j $ and $ m_{j} $ in the present section. The  general basis set describing both the vibration and rotation of the triatomic molecule is written:
\begin{equation}
	\ket{n_1n_2n j \bar{k}m_{j},p} = \ket{n_1}\ket{n_2}\ket{n j\bar{k}m_{j},p}.
\end{equation} 
where $\ket{nj\bar{k}m_{j},p}$ 
is the symmetrized rovibrational basis set describing the bending and rotation part as defined by Sutcliffe and Tennyson \cite{Sutcliffe-Mol-Phys-1986} and also used  in our previous work \cite{H2-H2O-Stoecklin-2019} :

\begin{equation}
	\ket{jn\bar{k}m_{j},p} = \dfrac{1}{\sqrt{2(1+\delta_{\bar{k},0})}}\left[\ket{jn\bar{k}m_{j}} + (-1)^p \ket{jn -\bar{k}m_{j}}\right]		
\end{equation}
with
\begin{equation} 
	\ket{jn\bar{k}m_{j}} = P^{\bar{k}}_{n}(cos \alpha )\ket{j\bar{k}m_{j}}
\end{equation}
where $m_{j}$  and $\bar{k}$ are the projections of the \cf{H2O}  rotational angular momentum $j$ along the  $z'$  space fixed and $z$ molecular fixed axis respectively. $P^{\bar{k}}_{n}(cos \alpha )$ is a generalized Legendre function while  $\ket{j\bar{k}m_{j}}$ is a symmetric top wave function. 
$\ket{n_1}$ and $\ket{n_2}$ are the eigenvectors of the radial
reference Hamiltonians $\hat{h}_i, i=1,2$ in a PODVR approach
\cite{H2O-Cl-Guo-2021,Optimized-DVR-Clary-1992}:
\begin{equation}
	\hat{h}_i\ket{n_i} = \varepsilon^i_{n_i}\ket{n_i} , i=1,2.
\end{equation}
We use a Gauss-Lanczos-Morse-DVR (GLM-DVR) basis \cite{Szalay1993} as a primitive basis for the PODVR, instead of the usual sine-DVR basis
considered for example in Ref.[~\citenum{H2O-Cl-Guo-2021}] as it describes
better the anharmonicity of the radial reference potentials. The
resulting irregular radial grid gives a very good accuracy for a
reduced number of points.

For A$_2$B molecules one may furthermore take advantage of the permutation symmetry and use a symmetrized radial basis:
\begin{equation}
	\ket{n_1 n_2,q} = N_{n_1,n_2}\left[\ket{n_1n_2}+(-1)^{q+\bar{k}}\ket{n_2n_1}\right] 
\end{equation}
with $N_{n_1,n_2} = \dfrac{1}{\sqrt{2(1+\delta_{n_1,n_2})}}$, $n_1 \ge
n_2$ for $q+\bar{k}$ even and $n_1 > n_2$ for $q+\bar{k}$ odd
\cite{DVR3D-TENNYSON-2004}. This expression, as discussed in Tennyson
\textit{et~al.} \cite{DVR3D-TENNYSON-2004}, presents the great
advantage to allow selecting \text{ortho} and \textit{para} symmetry
blocks of A$_2$B molecules by choosing a value of $q$. In the case of
water $q=0$ and $q=1$ are respectively associated with the
\textit{para} and \textit{ortho} symmetries.

From the matrix elements of the triatomic hamiltonian in the
unsymmetrized basis set reported in Refs.[~\citenum{Wang2017, Yang2021}]
one can easily derive the non-zero matrix elements in the symmetrized
basis:

\begin{eqnarray}
	\begin{array}{l}
		\bra{n_1n_2n j\bar{k}m_{j},q,p}\hat{h}_{1} +  \hat{h}_{2}\ket{n'_1 n'_2 n j \bar{k}m_{j},q,p}=
		2N_{n_1,n_2}N_{n_1',n_2'}(\varepsilon_{n_1}+\varepsilon_{n_2})\left[\delta_{n_1,n_1'}\delta_{n_2,n_2'} + (-1)^{q+\bar{k}}\delta_{n_1,n_2'}\delta_{n_2,n_1'}\right] \\
	\end{array}
\end{eqnarray}

\begin{eqnarray}
	\begin{array}{l}
		\bra{n_1 n_2 n j \bar{k}m_{j},q,p}\hat{T}_{\alpha} +  \hat{T}_{vr}\ket{n'_1 n'_2 n'  j\bar{k}m_{j},q,p}= 
		M_{n_1,n_2,n_1',n_2',q,\bar{k}} \times
		\left\{\dfrac{1}{8}\left[j(j+1)-\bar{k}^2\right]\delta_{n,n'}  + n(n+1)\delta_{n,n'} + \dfrac{1}{4}\left[j(j+1)-3\bar{k}^2\right]E_{n,n'\bar{k}} \right\}
	\end{array}
\end{eqnarray}

\begin{eqnarray}
	\begin{array}{l}
		\bra{n_1 n_2 n j \bar{k}m_{j},q,p}\hat{T}_{\alpha} +  \hat{T}_{vr}\ket{n'_1 n'_2 n'  j\bar{k}\pm 1 m_{j},q,p}=\dfrac{1}{4}(1+\delta_{\bar{k},0})^{\frac{1}{2}}(1+\delta_{\bar{k}\pm 1,0})^{\frac{1}{2}}\lambda^{\pm}_{j\bar{k}}R_{n_1,n_2,n_1',n_2',q,\bar{k}}\\ \times\left\{(2\bar{k}+1)(G_{n,n',\bar{k}}-D_{n,n',\bar{k}})-2\lambda^{\pm}_{n,\bar{k}}\delta_{n,n'} \right\}
	\end{array}
\end{eqnarray}

\begin{eqnarray}
	\begin{array}{l}
		\bra{n_1 n_2 n j \bar{k}m_{j},q,p} \hat{T}_{\alpha} +  \hat{T}_{vr} \ket{n'_1 n'_2 n'  j\bar{k}\pm 2 m_{j},q,p}=
		\dfrac{1}{16}(1+\delta_{\bar{k},0})^{\frac{1}{2}}(1+\delta_{\bar{k}\pm 2,0})^{\frac{1}{2}}\lambda^{\pm}_{j\bar{k}}\lambda^{\pm}_{j\bar{k}\pm 1}\times M_{n_1,n_2,n_1',n_2',q,\bar{k}}\times\left(2F_{n,n',\bar{k}}-H_{n,n',\bar{k}}\right)
	\end{array}
\end{eqnarray}
with $\lambda^{\pm}_{lm} = \sqrt{l(l+1)-m(m\pm 1)}$ and:
\begin{eqnarray}
	\begin{array}{l}
		M_{n_1,n_2,n_1',n_2',q,\bar{k}} = 2N_{n_1,n_2}N_{n_1',n_2'}\times\left[A_{n_1,n_2,n_1',n_2'} +(-1)^{q+\bar{k}} A_{n_1,n_2,n_2',n_1'}\right],
	\end{array}
\end{eqnarray}
\begin{eqnarray}
	\begin{array}{l}
		R_{n_1,n_2,n_1',n_2',q,\bar{k}} = 2N_{n_1,n_2}N_{n_1',n_2'}\times\left[B_{n_1,n_2,n_1',n_2'} +(-1)^{q+\bar{k}} B_{n_1,n_2,n_2',n_1'}\right],
	\end{array}
\end{eqnarray}

\begin{equation}
	A_{n_1,n_2,n_1',n_2'} = \bra{n_1n_2}\dfrac{1}{2\mu_1 R_1^2}+\dfrac{1}{2\mu_2 R_2^2}\ket{n_1'n_2'},
\end{equation}
\begin{equation}
	B_{n_1,n_2,n_1',n_2'} = \bra{n_1n_2}\dfrac{1}{2\mu_1 R_1^2}-\dfrac{1}{2\mu_2 R_2^2}\ket{n_1'n_2'}.
\end{equation}
where the $E_{n,n',\bar{k}}$, $G_{n,n',\bar{k}}$, $D_{n,n',\bar{k}}$,
$F_{n,n',\bar{k}}$ and $H_{n,n',\bar{k}}$ are angular integrals
defined in Ref.[~\citenum{Wang2017}] which are computed using a
Gauss-Legendre quadrature. For $\bar{k}=\bar{k}'=1$ there is an extra
term in the symmetrized hamiltonian coming from the
$\bar{k}\pm2,\bar{k}$ elements when $\bar{k}'=-1$ and $\bar{k}=1$ in
the unsymmetrized basis. This extra terms is defined as:

\begin{eqnarray}
	\begin{array}{l}
		\bra{n_1 n_2 n j (\bar{k}=1)m_{j},q,p} \hat{T}_{\alpha} +  \hat{T}_{vr} \ket{n'_1 n'_2 n'  j(\bar{k}=1) m_{j},q,p}=\dfrac{(-1)^{1-p}}{16}j(j+1)M_{n_1,n_2,n_1',n_2',q,\bar{k}}\times
		\left(2F_{n,n',\bar{k}}-H_{n,n',\bar{k}}\right)
	\end{array}
\end{eqnarray}

The symmetrised integrals over the potential take the following form:

\begin{eqnarray}
	\begin{array}{l}
		\bra{n_1 n_2 n j \bar{k}m_{j},q,p} V_{res}(R_1,R_2,\alpha)\ket{n'_1 n'_2 n' j \bar{k}m_{j},q,p}=2 N_{n_1,n_2} N_{n_1',n_2'}\times\\
		\left[ \bra{n_1n_2 n j \bar{k}m_{j}}V_{res}(R_1,R_2,\alpha)\ket{n_1'n_2' n' j\bar{k}m_{j}}	 +(-1)^{q+\bar{k}}\bra{n_1n_2 n j\bar{k}m_{j}}V_{res}(R_1,R_2,\alpha)\ket{n_2'n_1' n' j \bar{k}m_{j}}
		\right]
	\end{array}
\end{eqnarray}
They are computed using a Gauss-Legendre quadrature for the
integration over the Radau bending angle ($\alpha$) and a GLM-PODVR
quadrature over the radial Radau coordinates.

\subsection{Close Coupling equations}\label{cc}
We follow closely the approach detailed in the original paper of
Phillips \textit{et al.} \cite{H2-H2O-General-CS-Green-1995} dedicated
to the rigid rotor approach of the \ce{H2-H2O } collisions where the
reader can find the details of the derivation. We here only give their
slightly modified form in order to refer to the three modes of
vibration of the \ce{H2O} molecule which can be coupled by the
intermolecular potential.  We do not include the \ce{H2} vibration in
our notation as it is neglected in the present approach. We will for
rotation refer to the angular momenta $ \overrightarrow{j_{1}}$,
$\overrightarrow{j_{2}} $ and $\overrightarrow{l} $ associated
respectively with \ce{H2O}, \ce{H2} and their relative angular
momentum while $\overrightarrow{j_{12}} = \overrightarrow{j_{1}} +
\overrightarrow{j_{2}}$ and the $ \tau $ index allows distinguishing the \ce{H2O} rotational states inside a given j$ _{1} $ multiplet. In order to further simplify the notations we
will use the collective index $ ( i) $ to designate the set of quantum
numbers ($\nu_{1}\nu_{2}\nu_{3} j_{1} \tau j_{2} $ ) associated with a
given internal state of the \ce{H2-H2O} complex, where
$(\nu_{1}\nu_{2}\nu_{3})$ are the vibrational quantum numbers for,
respectively, the symmetry stretching, the bending, and the
antisymmetric stretching of \ce{H2O}. The close-coupling equations for
the radial part of the scattering wave function read:
\begin{eqnarray}
	\begin{array}{l}\label{cc-eq}
		\left\{\dfrac{d^{2}}{dR^{2}}-\dfrac{l(l+1)}{R^{2}}+k^{2}_{(i)}\delta_{(i),(j)}\delta_{j_{12}j_{12}'}\delta_{ll'} -[U(R)]_{ (i) j_{12} l; (j)  j'_{12} l' }^{JMP} \right\}  
		\times G_{(i)  j_{12} l ; (j)  j'_{12} l'} ^{JMP}(R)=0
	\end{array}
\end{eqnarray}
where $  G_{(i) j_{12} l; (j) j'_{12} l'}^{JMP}(R) $ is the radial part of the \ce{H2-H2O} scattering wave function,  $ k^{2}_{(i) }=2\mu [E- \varepsilon_{(i)} ] $ and 
\begin{eqnarray}
	\begin{array}{l}\label{eqn:pot}
		[U(R)]_{(i) j_{12} l ; (j) j'_{12}l'}^{JMP} = 2\mu \sum_{p_{1} q_{1}p_{2} p}	
		\langle\, (i) j_{12} l \,J M|t_{p_{1} p_{2} p}^{q_{1}}| (j) j'_{12}l' \, J M \,\rangle  \times\\
		\int\limits_{0}^{\pi} d\alpha\sin(\alpha)\int dR_{1} \int dR_{2}[{\varGamma}_{\nu_{1} \nu_{2} \nu_{3}} ^{j_{1}\tau}  (R_{1},R_{2},\alpha)\times
		V_{p_{1} p_{2} p}^{q_{1}}(R_{1},R_{2},\alpha,R)\times
		{\varGamma}_{\nu'_{1} \nu'_{2} \nu'_{3}} ^{j'_{1}\tau'}  (R_{1},R_{2},\alpha)]
		
	\end{array}
\end{eqnarray}
In this expression the ${\varGamma}_{\nu_{1} \nu_{2} \nu_{3}}
^{j_{1}\tau} (R_{1},R_{2},\alpha)$ are the \ce{H2O} wave functions
calculated in the previous section while the \\$ \langle\, (i) j_{12}
l \,J M|t_{p_{1} p_{2} p}^{q_{1}}| (j) j'_{12}l' \, J M \,\rangle $
are the rigid rotor analytical angular matrix elements of the
potential given in the paper of Phillips \textit{et~al.}
\cite{H2-H2O-General-CS-Green-1995}. $J$ and $M$ are respectively the
total angular momentum and its projection along the Z space fixed axis
while $P$ is the parity of the system. We furthermore replace the $ \varepsilon_{(i)}  $ calculated in the previous section by their experimental values.

A new code called \texttt{Divitas} and including many of the
ingredients of the \texttt{Newmat} \cite{Newmat:02} and
\texttt{Didimat}\cite{H2-CO-scientific-report} code was written in
Bordeaux. It includes the calculations of the ro-vibrational states of
\ce{H2O } detailed in section \nameref*{h2o-bs} and solves the space
fixed Close Coupling equations (\ref{cc-eq}) using a log derivative
propagator \cite{Manolopoulos1986}.  It was tested by reproducing the
rigid rotor calculations performed in Grenoble using the
\texttt{Molscat} code \cite{molscat2012}.

\section{Calculations and results}\label{CalRes} 

\subsection{H$ _{2} $O bound states} 

We use the most recent POKAZATEL PES of water \cite{Polyansky2018} to
compute the energy levels of the water molecule for 8 values of the
rotational state $j_1=0-7$. This PES which is the most accurate available was specifically designed to describe very excited ro-vibrational \ce{H2O} states.  Our basis set contains 4 GLM-PODVR functions for each $R_1$ and $R_2$ radial coordinates 
and 45 values of $n$. Angular
integrals were computed using 50 Gauss-Legendre quadrature points
while the radial integrals use a two dimensional $R_1,R_2$ grid of 10
GLM-PODVR points for $q+\bar{k}$ even and 6 points for $q+\bar{k}$
odd due to the use of radial symmetrized basis.

The GLM-PODVR basis set is eigenfunction of the 1D reference
hamiltonian $\hat{h}_1$ (due to the symmetry of the water molecule
$\hat{h}_1$ and $\hat{h}_2$ are identical ) and is obtained using a
100 point 1D GLM-DVR spanning the [0.9,4.5] $a_o$ interval. A grid
of 30 points is enough to achieve a convergence better than 0.3
cm$^{-1}$ for all the calculated states and better than 0.01 cm$^{-1}$
if we consider only the rotational states inside the $(0,0,0),(0,1,0)$
and $(0,2,0)$ vibrational states.  The comparison with experiment
shows the very good accuracy achieved in the present calculations. For
all the 320 ro-vibrational states computed in the present work, the
maximum percentage error was 0.222 \% for \textit{para} states and
0.218 \% for the \textit{ortho} ones.

\begin{table}
	\centering
	\begin{tabular}{cccc}
		\hline
		\hline
		\\
		($\nu_1$,$\nu_2$,$\nu_3$)&(EXP-TW) & (EXP-TW2)&(EXP-POL)\\
		\hline
		(0,1,0) & 0.374 & 0.158& -0.310 \\
		(0,2,0) & 0.727 & 0.484& -0.540 \\
		(1,0,0) & 0.840 &-0.242 & -0.840 \\
		(0,0,1) & 0.838 &-0.039 & -0.730 \\
		\hline	
	\end{tabular}
	\caption{Comparison between theory and experiment of the 4 first excited vibrational energy levels of \ce{H2O} ($j=0$) calculated using our approach and the POKAZATEL PES\cite{Polyansky2018} (TW) or the PES reported in Ref~\citenum{Polyansky2013} (TW2). The errors reported in Ref~\citenum{Polyansky2013} (POL) using DVR3D are reported in the last column. Units are cm$ ^{-1} $.} \label{table2}
\end{table} 

\begin{table*}\label{table1}
	\begin{tabular}{ccccccccccc}
		\hline
		\hline
		\\
		($\nu_1$,$\nu_2$,$\nu_3$)&$j_{kakc}$ & EXP &(EXP-YA) &(EXP-TW)& &($\nu_1$,$\nu_2$,$\nu_3$)&$j_{kakc}$ & EXP & (EXP-YA) &(EXP-TW)\\
		\\
		\hline
		para-H$_2$O		&		&&&&				&ortho-H$_2$O\\	
		(0,0,0)&$0_{00}$ & 0 	    & 0 	 & 0 & &(0,0,0)&$1_{01}$ & 23.794 	& 0.094  & -0.016 \\
		(0,1,0)&$0_{00}$ & 1594.746 & -0.754 & 0.374 &&(0,1,0)&$1_{01}$ & 1618.557 & -0.643 & 0.302  \\  \\
		(0,2,0)&$0_{00}$ & 3151.630 & -2.170 & 0.727  &&(0,2,0)&$1_{01}$ & 3175.441 & -2.159 & 0.598  \\
		(1,0,0)&$0_{00}$ & 3657.053 & 9.253	 & 0.840 &&  (1,0,0)&$1_{01}$ & 3680.454 & 9.354	 &0.877	 	\\
		&$1_{11}$ & 3693.294	&9.494	 &0.857    && &$1_{10}$ & 3698.491 & 9.491	 &0.957	 \\
		&$2_{02}$ & 3725.942	&9.442	 &0.920		&& &$2_{12}$ & 3734.897	& 9.597	 &0.837   \\
		&$2_{11}$ & 3750.465	&9.565	 &0.934		&& &$2_{21}$ & 3788.694	& 10.094 &1.079		\\ 
		&$2_{20}$ & 3789.969	&10.069	 &0.882		&& &$3_{03}$ & 3791.372	& 9.672	 &0.887		\\ 
		&$3_{13}$ & 3796.540	&9.740	 &0.980		&& &$3_{12}$ & 3827.393	& 9.793	 &1.134		\\
		&$4_{22}$ & 3966.559	&10.359	 &0.911		&& &$3_{21}$ & 3864.764 & 10.164 &0.961	\\
		&$5_{33}$ & 4150.287 &11.487	 &1.388		&& &$3_{30}$ & 3935.345	& 11.045 &1.014  \\
		(0,0,1)&$1_{01}$ & 3779.493	&11.193	 &0.783		&& (0,0,1)&$0_{00}$ & 3755.928	& 11.128 &0.838	\\
		&$3_{12}$ & 3926.862	&11.662	 &1.218		&& &$3_{22}$ & 3956.666 & 12.066 &1.337       \\
		\\
		\hline
	\end{tabular}
	\caption{Comparison of the experimental energy levels of
		\textit{para}- and \textit{ortho}-H$_2$O (EXP) with the theoretical
		results reported in Yang \textit{et~al.}\cite{Guo_ar_2022} (YA) and
		the ones obtained in this work (TW). The experimental values as well
		as the absolute errors are reported in $cm^{-1}$. \label{po-states}}
\end{table*}
Very accurate \cf{H2O} ro-vibrational bound state energies were reported by POKAZATEL \cite{Polyansky2018}. These authors used the parametrised POKAZATEL PES\cite{Polyansky2018} that was calculated from \textit{ab initio} points and fitted to the experimental energy values with the help of the \texttt{DVR3D} package \cite{DVR3D-TENNYSON-2004}. This semi empirical PES  error leads to a very  good accuracy of the bound state energies which is better than 0.05 cm$^{-1}$ provided that the DVR3D method is used for the bound states calculations as it was used to perform the fit. As our PES is not fitted on experiment, we compare our results with other non-semi-empirical values provided by the same authors using just an \textit{ab initio} PES\cite{Polyansky2013}. We present in Table \ref{table2} the results of this comparison and observe their excellent agreement. We furthermore can see that our most accurate results are obtained from the PES of Ref.~\citenum{Polyansky2013} which is not semi empirical. We then conclude that the use of the \textit{ab initio} + semi-empirical PES\cite{Polyansky2018} leads to better results only if the DVR3D method  which was used to perform the fit is also used to calculate the levels. A comparison with the rovibrational states recently computed by Yang \textit{et al.}
\cite{Guo_ar_2022} using their \textit{ab initio} PES is also presented in Table \ref{po-states}. The
present approach appears to be more accurate by about a factor of ten.

\subsection{Close-Coupling calculations}

As mentioned in section \ref{cc} the new code is an extension of the
\texttt{Newmat} code version which was recently developed for treating the
bending relaxation of \ce{H2O} colliding with an atom
\cite{H2-H2O-Stoecklin-2019}. It was then tested by reproducing
calculations performed in recent work for the bending relaxation of
\ce{H2O} by collisions with He \cite{Stoecklin-H2o-He-2021}. It was
also tested by reproducing Molscat rigid rotor calculations for the
\ce{H2 + H2O} collisions \cite{Dubernet-2012}using the experimental values of the \ce{H2} rotational constants \cite{Huber1979}. For all the
calculations presented, three levels of each \ce{H2} spin modification
were necessary to converge the bending relaxation cross
section. Rotational states of \ce{H2O} for $ j_{1} $=0 to 5 were
included in the two bending states considered in the
calculations. This choice of basis resulting from a compromise between
computer time and accuracy allowed us to include the $j=5$ level of
\ce{H2} which was neglected in previous calculations
\cite{H2-H2O-Wiesenfeld-2022}. We checked that increasing the
rotational basis set of \ce{H2O} up to j=7 only marginally affects the magnitude
of the bending relaxation cross section for levels with $j\leq 5$. A Chebyshev grid of 18
bending angles and the same two dimensional $R_1$,$R_2$
GLM-PODVR grid used before for the calculation of the wave
functions of \ce{H2O} was used to calculate the 
vibrational part of the potential matrix elements while 20
closed channels were included for each collision
energy in the 0.1-1000 cm$ ^{-1} $ interval. This is corresponding to
a maximum rotational energy of about 4200 cm$ ^{-1}$, which is
slightly higher than the 3500 cm$ ^{-1} $ used by Wiesenfeld
\cite{H2-H2O-2021-Wiesenfeld}. Relative convergence of the bending
relaxation cross section was tested as a function of the total angular
momentum $J$ for each collision energy to be better than $ 10^{-6} $
leading to a maximum value of $J=65$ at 1000 cm$ ^{-1}$, which is more
than the double of the maximum value of $J=32$ used by Wiesenfeld
\cite{H2-H2O-Wiesenfeld-2022}.

\begin{figure} 
	\centering
	\includegraphics[width=0.5\textwidth]{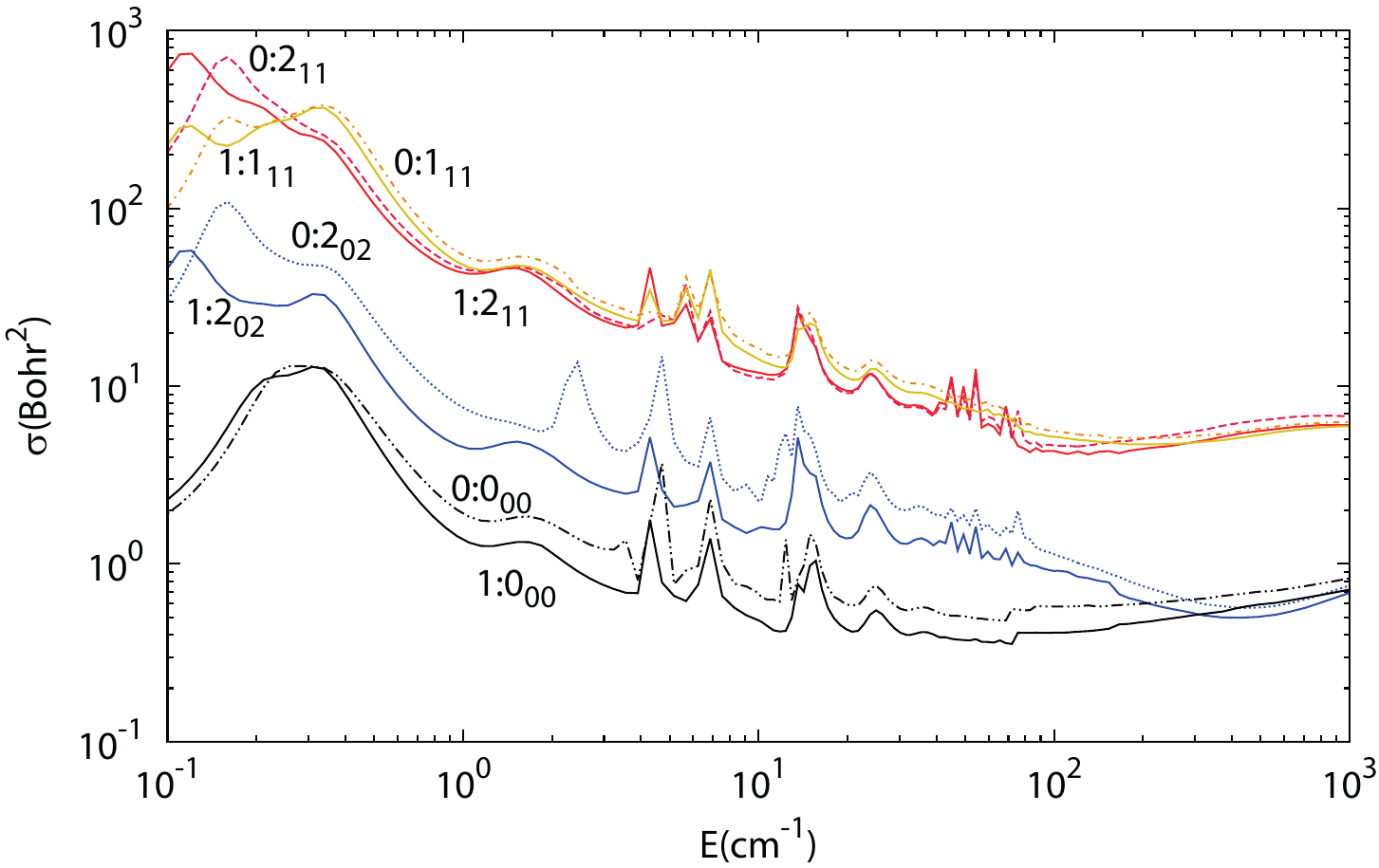}
	\caption{Comparison between rotationally inelastic state to
		state cross section for \ce{H2O}$ (\nu_{2}, 2_{20}) $
		colliding with \textit{para}-H$ _{2}(j=0) $ inside $
		\nu_{2}= $ 0 and 1. The final \ce{H2O} ro-vibrational state is
		indicated on each curve. The solid and dashed line are
		respectively associated with transitions inside $ \nu_{2}= $
		1 and 0.  \label{pH2-pH2O-compar}}.
\end{figure}
\begin{figure}
	\centering
	\includegraphics[width=0.5\textwidth]{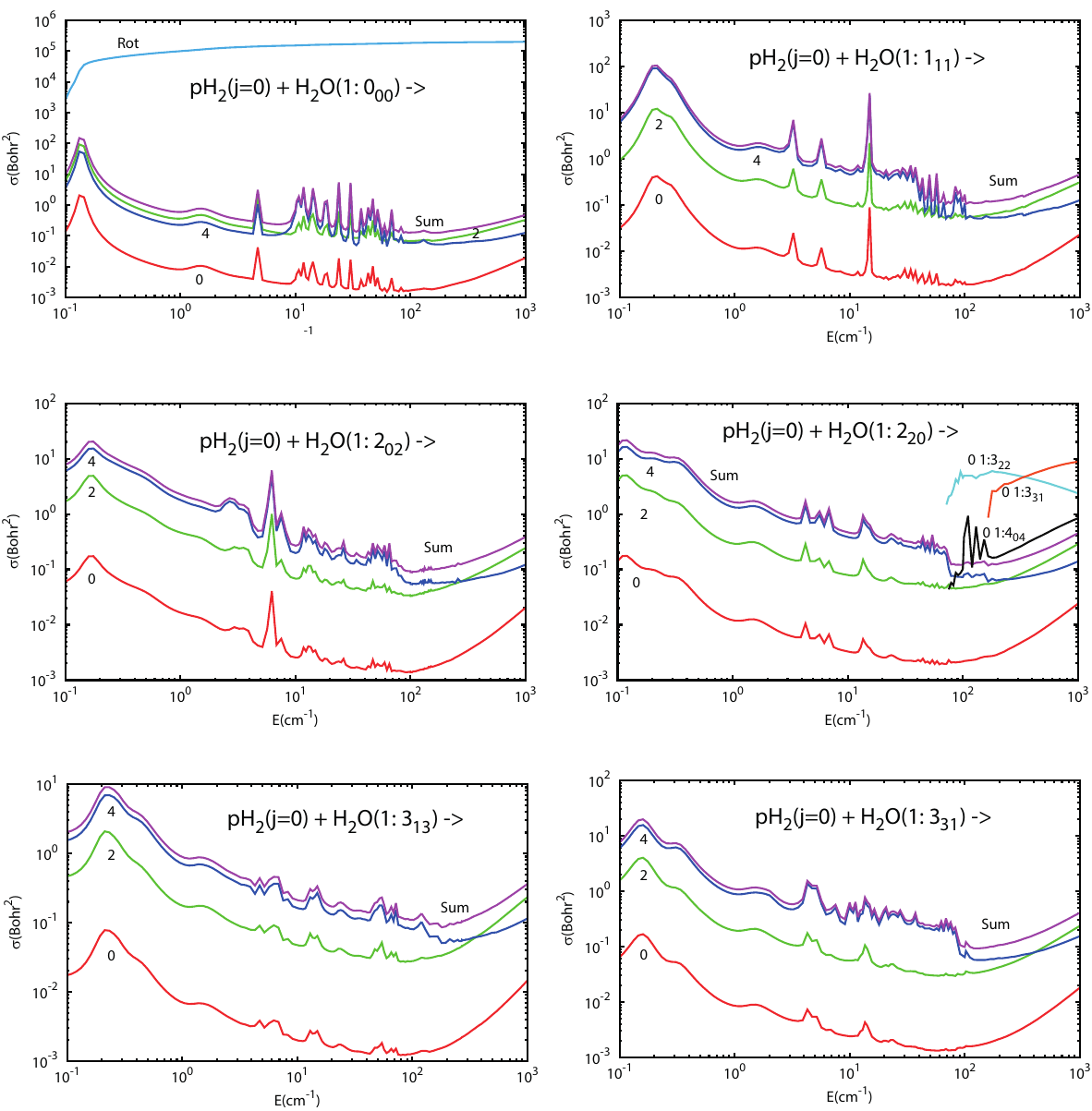}
	\caption{Comparison between the vibrational quenching cross
		section of \textit{para}-\ce{H2O}$ (\nu_{2}=1,
		j_{k_{A}k_{C}}) $ by collision with \textit{para}-H$
		_{2}(j=0) $ for several initial levels of \ce{H2O} belonging
		to the $\nu_{2}=1, j=0, 1, 2$ and $3$ manifolds. The curves 0, 2, 4 correspond to the final state of \ce{H2}. In the middle right panel the opening of  excited state of \ce{H2O} is shown. The first reported quantum number on these curves is denoting the rotation of \ce{H2} while the others give the \ce{H2O} ro-vibrational state. In light blue (upper left panel) the summed rotationally inelastic cross section inside the  initial $\nu_{2}=1$  level is also represented for comparison.  \label{pH2-pH2O}}
\end{figure}
\begin{figure}
	\centering
	\includegraphics[width=0.5\textwidth]{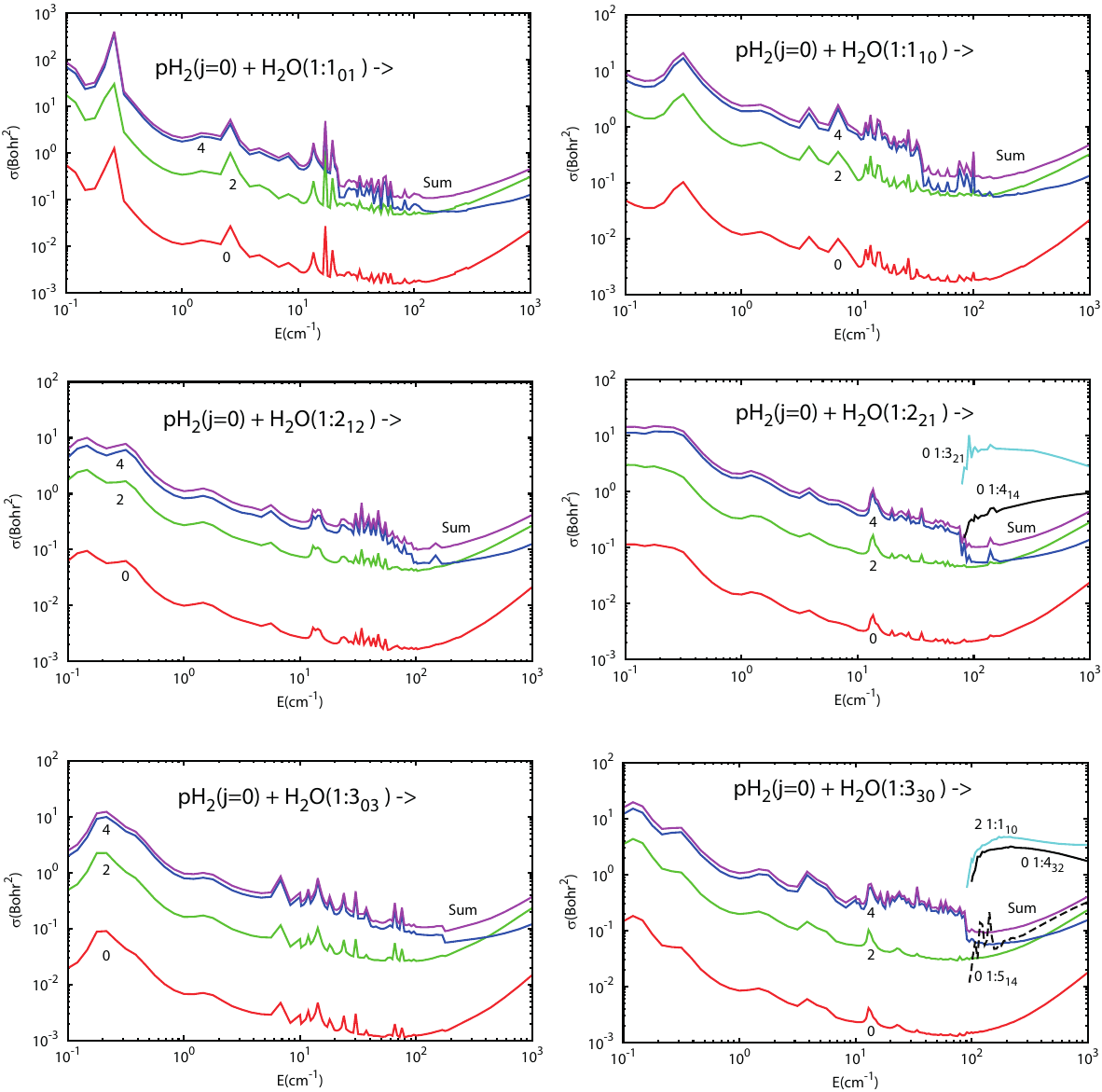} 
	\caption{Comparison between the vibrational quenching cross
		section of \textit{ortho}-\ce{H2O}$ (\nu_{2}=1,
		j_{k_{A}k_{C}}) $ by collision with \textit{para}-H$ _{2}
		(j=0)$ for several initial levels of \ce{H2O} belonging to
		the $ \nu_{2}=1, j=1, 2$ and $3$ manifolds. In the lower and middle right panels the opening of excited states (higher in energy than the initial state ) of \ce{H2O} is also shown.\label{pH2-oH2O}  The first reported quantum number on these curves is denoting the rotation of \ce{H2} while the others give the \ce{H2O} ro-vibrational state.}
\end{figure}

\begin{figure}
	\centering
	\includegraphics[width=0.5\textwidth]{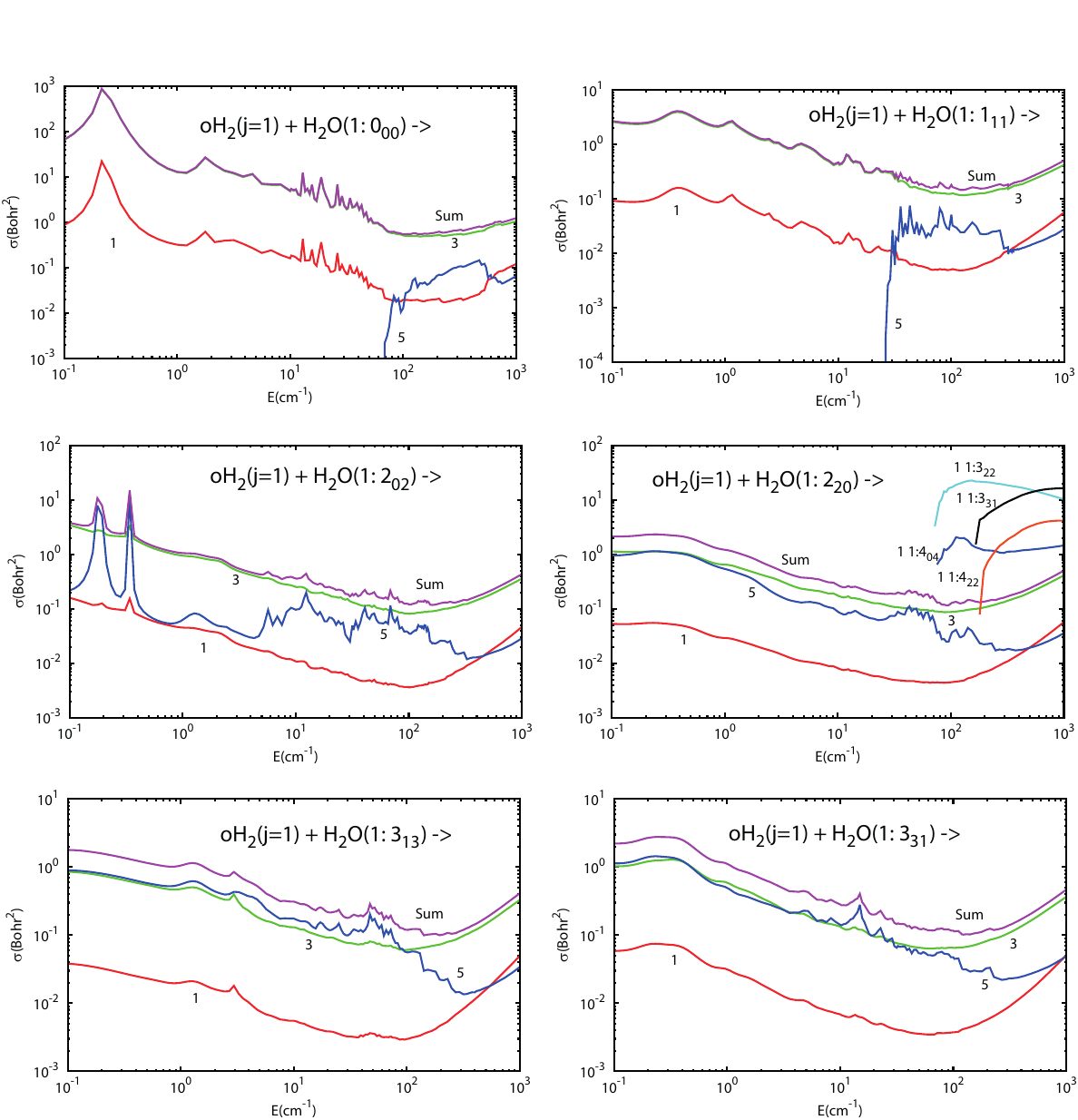} 
	\caption{Comparison between the vibrational quenching cross
		section of \textit{para}-\ce{H2O}$ (\nu_{2}=1,
		j_{k_{A}k_{C}}) $ by collision with \textit{ortho}-H$ _{2}
		(j=1)$ for several initial levels of \ce{H2O} belonging to
		the $ \nu_{2}=1, j=0, 1, 2$ and $3$ manifolds. The curves 1, 3, 5 correspond to the final state of \ce{H2}. In the middle right panel, state to state cross sections associated with the opening of a few excited  levels (higher in energy than the initial state ) of \ce{H2O} are also represented. The first reported quantum number on these curves is denoting the rotation of \ce{H2} while the others give the \ce{H2O} ro-vibrational state.\label{oH2-pH2O}}
\end{figure}
\begin{figure}
	\centering
	\includegraphics[width=0.5\textwidth]{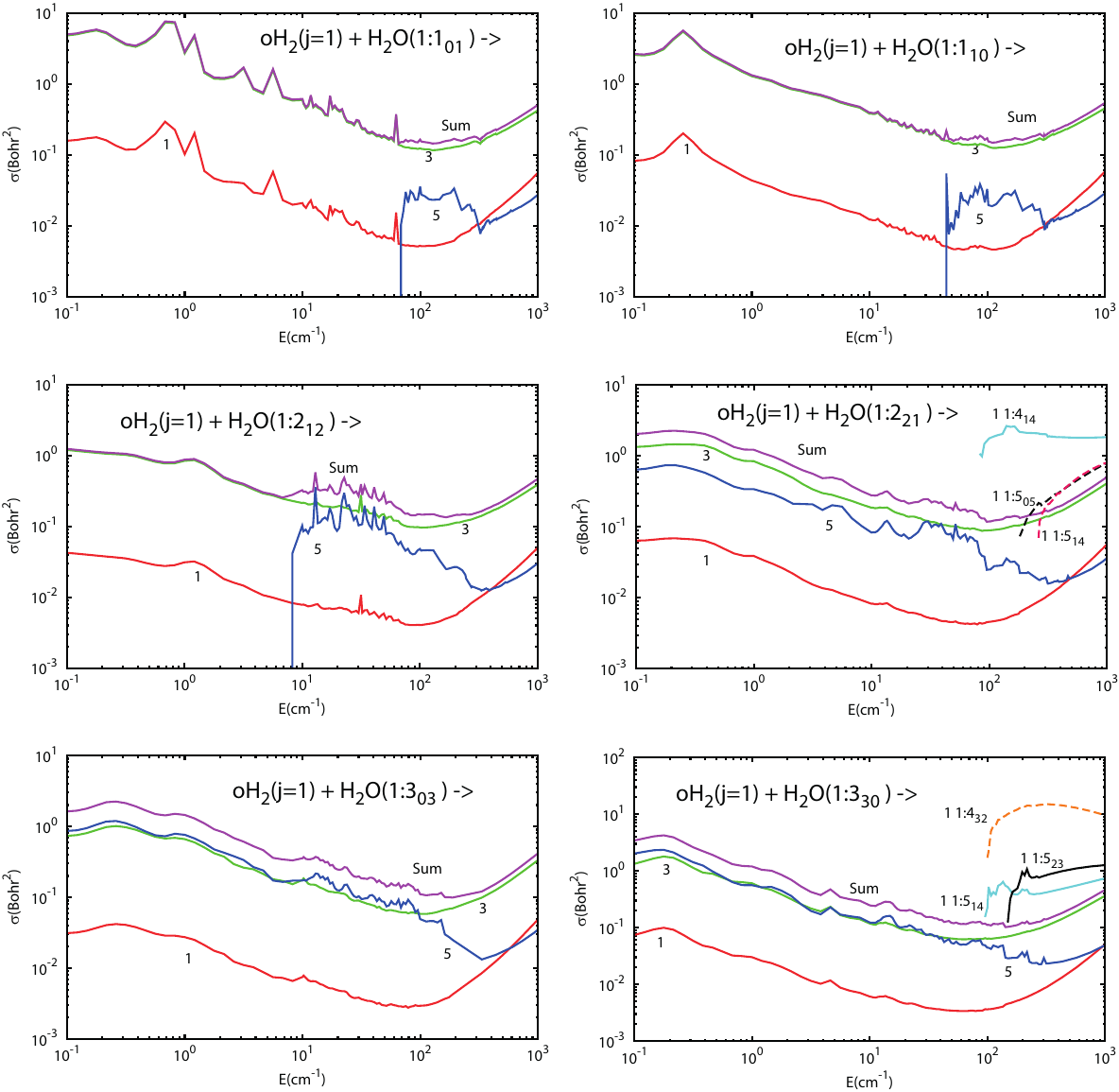} 
	\caption{Comparison between the vibrational quenching cross
		section of \textit{ortho}-\ce{H2O}$ (\nu_{2}=1,
		j_{k_{A}k_{C}}) $ by collision with \textit{ortho}-H$ _{2}
		(j=1)$ for several initial levels of \ce{H2O} belonging to
		the $ \nu_{2}=1, j=1, 2$ and $3$ manifolds.   In  the middle and lower right  panels the opening of excited  states (higher in energy than the initial state ) of \ce{H2O} is also shown.  The first reported quantum number on these curves is denoting the rotation of \ce{H2} while the others give the \ce{H2O} ro-vibrational state.	\label{oH2-oH2O}}
\end{figure}

\begin{figure}
	\centering
	\includegraphics[width=0.4\textwidth]{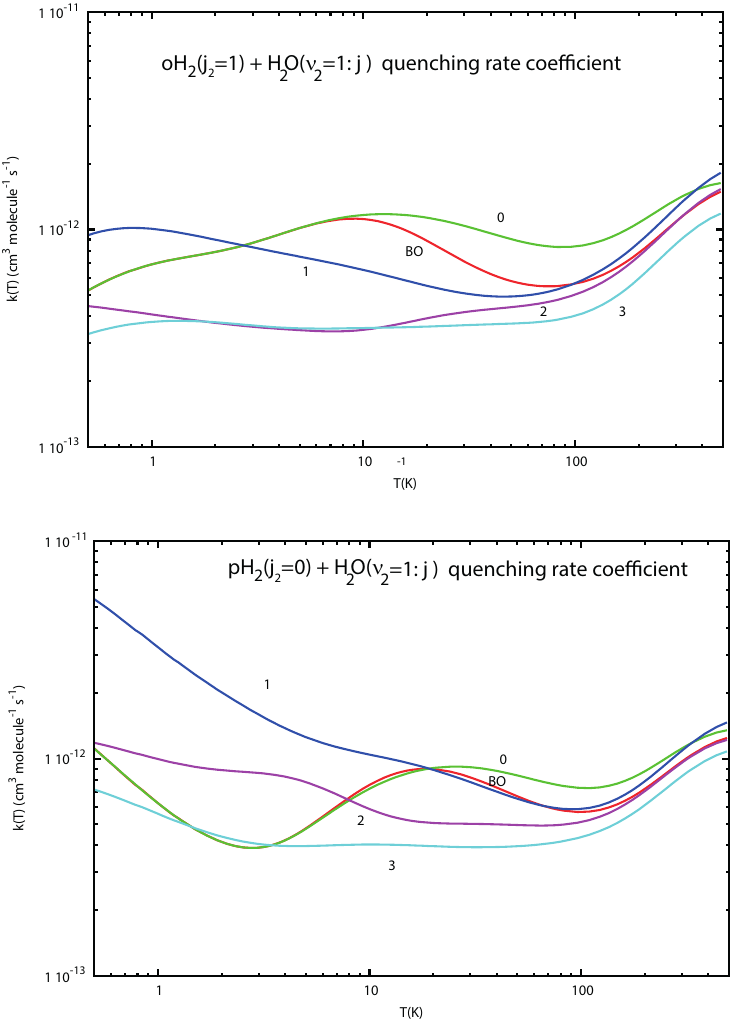}
	\caption{Comparison between the vibrational quenching rate
		coefficients of the $ (\nu_{2}=1, j=0, 1, 2 $ and $3$)
		states of \ce{H2O} by collision with \ce{H2} and the
		Boltzmann averaged rate coefficient. The collisions with
		\textit{para}-\ce{H2}($j_2=0$) and with
		\textit{ortho}-\ce{H2} ($j_2=1$) are respectively presented
		in the lower and upper panels. \label{H2-H2O-j-ba}}
\end{figure}

\begin{figure}
	\centering
	\includegraphics[width=0.5\textwidth]{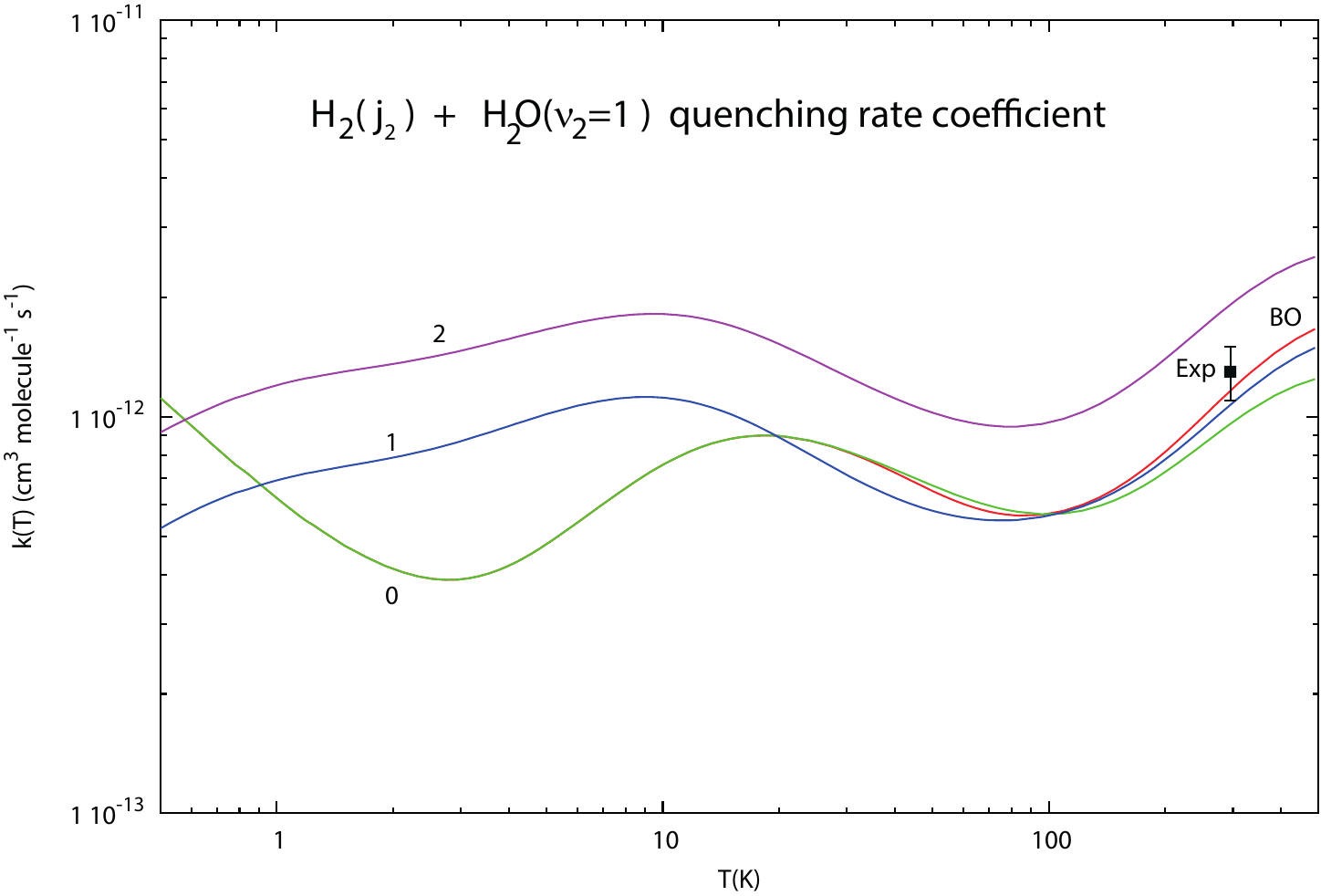} 
	\caption{Comparison between the vibrational quenching rate
		coefficients of \ce{H2O} $ (\nu_{2}=1)$ by collision with
		the two spin modifications of \ce{H2} and the global
		Boltzmann averaged rate coefficient. The only experimental
		value\cite{H2O-H2-quenvib-exp-1991} available at 295 K is
		also reported. \label{global-rate}}
\end{figure}

\subsubsection{Rotational transitions inside a given bending level of \ce{H2O}}

We compare in Figure \ref{pH2-pH2O-compar} pure rotational transitions
inside $ \nu_{2}= $ 0 and 1 and starting from the $ 2_{20} $ level of
\ce{H2O} in collision with \textit{para}-H$ _{2}(j_2=0)$.  The final state of \ce{H2} is   $ (j_2=0) $ for all the \ce{H2O} transitions represented.  As can be seen
on this figure the position of the resonances are about the same for
the transitions considered inside the two different bending
levels. The magnitudes are conversely seen to be slightly lower in the
excited $ \nu_{2}= $ 1 level. This is not surprising as channels
associated with excited levels of \ce{H2} (for example $j_2$=4 +
\ce{H2O}($ \nu_{2}= 0, 5_{24} )$ ) are open and interspersed between
open $ \nu_{2}= $ 1 levels. This means that part of the scattering
flux is directed to this open levels, lowering the magnitude of the
cross sections obtained for the same transition inside the $ \nu_{2}=
$ 0 bending level.

\subsubsection{Bending relaxation propensity rules}

\paragraph*{ Impact of the initial and final \ce{H2} rotational basis:}

Figures \ref{pH2-pH2O}, \ref{pH2-oH2O}, \ref{oH2-pH2O}, \ref{oH2-oH2O}
show the bending relaxation cross sections for the four combinations
of spin modifications of \ce{H2} and \ce{H2O} respectively
\textit{p-}\ce{H2} + \textit{p-}\ce{H2O} ; \textit{p-}\ce{H2} +
\textit{o-}\ce{H2O} ; \textit{o-}\ce{H2} + \textit{p-}\ce{H2O} and
\textit{o-}\ce{H2} + \textit{o-}\ce{H2O}. On each of this figure the
contributions of the three rotational levels of \ce{H2} considered in
the calculations are reported as well as the total bending relaxation cross
sections, again summed over the final states of water.  If we consider
first Figures \ref{pH2-pH2O} and \ref{pH2-oH2O} associated both with
collisions of \textit{p-}\ce{H2} we see that the three $j_2=0, 2$ and
$4$ final channels are open and give (except for the initial level \textit{p-}\ce{H2} (j$ _{2} $=0) +\ce{H2O}(1:0$ _{00} $)) contributions in ascending order, the
global bending relaxation cross section being always almost equal to the $j_2 =
4$ contribution, except above $\sim 200- 300$~cm$^{-1}$. This is in
contrast with rotational relaxation which favours the lowest $ \Delta
j $. As a matter of fact vibrational relaxation is conversely
controlled by a near-resonant energy transfer mechanism which means
that the largest contributions to the bending relaxation cross sections are
provided by the internal levels of the complex which are the closest
in energy with the initial level. Pure rotational transitions inside
the excited bending level of \ce{H2O} give cross sections increasing
with collision energy and up to five orders of magnitude larger than
the bending relaxation cross sections, as illustrated in the upper left
panel of Figure \ref{pH2-pH2O}. This results in the steps
observed in the bending relaxation cross sections around 70 cm$ ^{-1} $
associated with the opening of rotational levels inside the $
\nu_{2}=1 $ excited bending level of \ce{H2O}, as shown in the middle right
panel of figure \ref{pH2-pH2O} and in the right  middle and lower panels of
figure \ref{pH2-oH2O} respectively for the opening of for example the \textit{p-}
\ce{H2}(j=0) + \ce{H2O}($ \nu_{2}=1,3_{22} $) and \textit{p-}
\ce{H2}(j=0)+ \ce{H2O}($ \nu_{2}=1,3_{21}$) resulting in a decrease of
the bending relaxation cross sections.

The strong dependence of the bending relaxation cross section as a
function of the size of the \ce{H2} rotational basis set included in
the calculation explains why our first estimate of the bending relaxation cross
section which considered only the $j=0$ state of \ce{H2} was two
orders of magnitude smaller than the experimental value at 300 K, as
suggested in that work\cite{H2-H2O-Stoecklin-2019}.

If we now compare Figures \ref{oH2-pH2O} and \ref{oH2-oH2O} associated
both with collisions of \textit{o-}\ce{H2} we see that the final level
$j_2 =5$ is closed at low collision energy for many initial
channels. This results in the main contribution to the bending
relaxation cross section to be due to the $j_2 =3$ final level for
almost all the initial levels represented. The contribution of the
$j_2 =5$ final level is furthermore seen to exhibit many resonances as
it is the closest in energy with the initial level and consequently
the most sensitive to the opening of rotational levels inside the $
\nu_{2}=1$ level of \ce{H2O}. \\

This important difference between the collisions of \textit{p-}\ce{H2}
and \textit{o-}\ce{H2} analysed in the previous paragraph results in dramatic differences between the
magnitude of the bending relaxation of \ce{H2O} at moderate collision
energy specially for the lowest rotational state of the excited $
\nu_{2}=1$ bending state. If we for example examine the case of the
collisions of the \textit{p-}\ce{H2O}( $\nu_{2}=1, 0_{00}$) state we
see that the rotational state closer in energy in the $ \nu_{2}=0$
manifold is the \textit{p-}\ce{H2}($j_2$=4) + \ce{H2O}($
\nu_{2}=0,4_{31} $) in the case of the collisions with
\textit{p-}\ce{H2} which is only about 24.5 cm$ ^{-1} $ lower in
energy than the initial state. In the case of the collisions with
\textit{o-}\ce{H2} it is the \textit{o-}\ce{H2}($j_2$=3) +
\textit{p-}\ce{H2O}($ \nu_{2}=0,5_{51}$) state which is about 259.5
cm$^{-1}$ lower in energy than the initial state. As bending
relaxation is controlled by a near-resonant energy transfer mechanism
this means that bending relaxation of water is stronger with \textit{p-}\ce{H2}($j_2=0$) than with
\textit{o-}\ce{H2}($j_2=1$) for moderate collision energies
corresponding to the region of the potential well which is 235 cm$ ^{-1} $ deep. 
There is, conversely,  no visible tendency to favour either the \textit{ortho} or the
\textit{para} transitions of \ce{H2O} for collisions with the same symmetry of \ce{H2}(\textit{o-} or  \textit{p-}\ce{H2})  as already noticed for  the quenching of \ce{H2O} by atoms. \cite{Stoecklin-H2o-He-2021}.

\paragraph*{ Impact of the initial \ce{H2O} rotational basis :}

$j$(\ce{H2O})-selected vibrational quenching rate coefficient of a
given $ (\nu_2=1,j) $ state of H$ _{2}$O are compared for the
collisions with the two spin modifications of \ce{H2} in their ground
state in Figure \ref*{H2-H2O-j-ba}. They were obtained by Boltzmann
averaging all the asymmetric top multiplet contributions (i.e. for
each $j$) in the temperature range $0.5-500$~K and by using the
\textit{ortho}:\textit{para} statistical weight 3:1 for the water
molecule. The global rate coefficient obtained by Boltzmann averaging
the $j=0-3$ selected rates is
also represented. \\ In the limit of low collision energies, the \ce{H2O} ($j=0$ ) vibrational quenching rate is seen to be increasing  as a
function of collision energy for the collisions with
\textit{o-}\ce{H2} while it  decreases as collision energy
increases for collisions with \textit{p-}\ce{H2}. We note that at higher
temperatures (above 1500 K) which were not explored in the present
approach, a rotational enhancement of the vibrational relaxation
process was observed at the classical level for low $\textit{j}$ by Faure
\textit{et~al.}  \cite{Valiron-2005}. We indeed can already see for
the collisions with both \textit{p-} and \textit{o-}\ce{H2} that the $j=1$ state selected
quenching rate becomes larger than the one of $j=0$ at higher
temperature. \\

We note that for astrophysical applications, the most extensive collisional dataset for H$_2$O + H$_2$ is currently that of Faure and Josselin \cite{Faure-vib-2008A&A}which includes all \ce{H2O} levels below 5000~cm$^{-1}$ and kinetic temperatures up to 5000~K. This dataset is based on literature data combined with a simple extrapolation procedure where rotation and vibration are fully decoupled. Our close-coupling calculations were limited here to water levels $j=0-5$ in the vibrational states (000) and (010) and kinetic temperatures lower than 500~K so that the present dataset as well as that of Wiesenfeld \cite{H2-H2O-Wiesenfeld-2022}  are too small for actual astrophysical applications. Calculations will be extended to higher \ce{H2O} ro-vibrational levels and higher kinetic temperatures in a future work and the new dataset will be made available in the BASECOL (https://basecol.vamdc.eu/) and EMAA databases (https://emaa.osug.fr/).

\paragraph*{ Comparison with experiment :}

The contributions of \textit{p-} and \textit{o-}\ce{H2} are Boltzmann
averaged over $j_2$=0,1,2 to
obtain the global thermal rate coefficient which is presented in
Figure \ref{global-rate}, where it is compared to the only
experimental data available at 295 K. A first interesting result
appears on this figure as contribution of \textit{o-} \ce{H2} is seen
to be negligible below 30 K, but slightly increases with temperature and becomes significant above 100K. Our computed value
at 295 K is $ 1.18 \times10^{-12} $ cm$ ^{3 } $molecule$ ^{-1}$s$
^{-1} $ in excellent agreement with  the experimental value of $ (1.3 \pm
0.2) \times10^{-12} $ cm$ ^{3 } $molecule$ ^{-1} $ s$ ^{-1} $.

Interestingly, our value is about three times lower than the
one obtained for \textit{o-}\ce{H2O} by Wiesenfeld
\cite{H2-H2O-2021-Wiesenfeld} who used a simple direct product of a
rigid rotor basis for the rotation and a normal mode function for the
vibration. It is however difficult to compare the two kind of calculations which differ in several other aspects. The rotational basis sets both for \ce{H2} and \ce{H2O} are different and  this author also performed  calculations only for selected values of J and interpolated the cross sections for the missing values of J. In order to analyse more clearly the effect of the bending-rotation coupling we then performed new calculations presented in the following paragraph.



\begin{figure}
	\centering
	\includegraphics[width=0.3\textwidth]{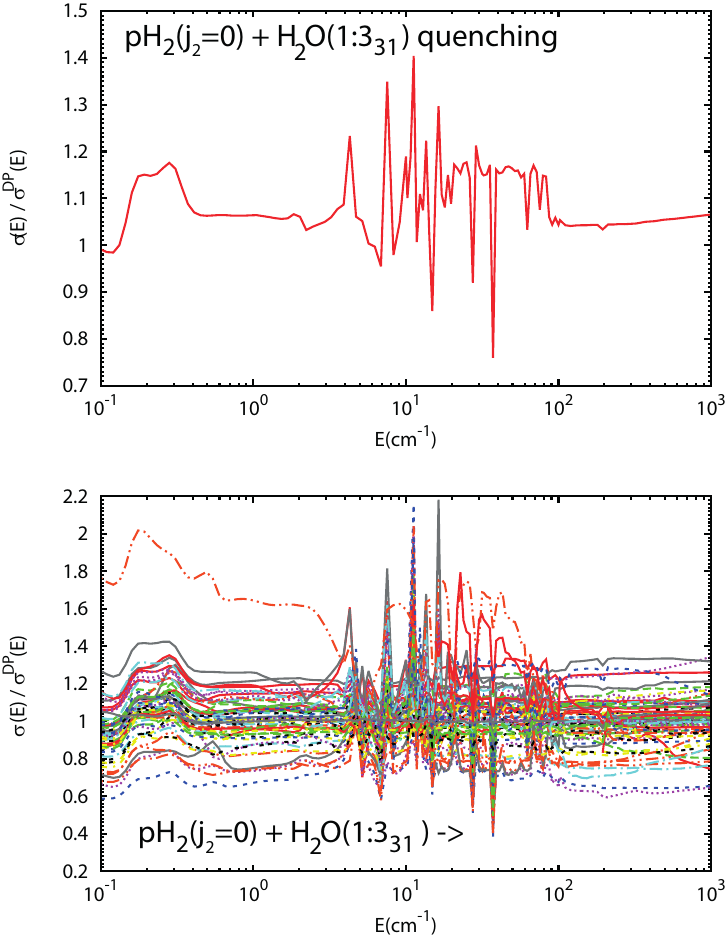} 
	\caption{This figure shows the ratio of Close Coupling and direct product state to state cross sections  as a function of collision energy for the collisions of  \textit{p-}\ce{H2} with  \ce{H2O}$ (\nu_{2}=1,3_{31}) $. \label{fig:ratio}}
\end{figure}

\paragraph*{ Bending-rotation coupling}
We investigate  the effect of the coupling between bending and rotation by performing a new set of calculations neglecting this coupling by using for water a direct product of mono dimensional bending wave function   and rigid asymmetric  top rotational wave functions while keeping the experimental ordering of the \ce{H2O} energies. As we are only interested, in the present study, in the two lowest bending levels of \ce{H2O} the coupling with the two stretches can be safely neglected. 
This approach is very similar to the one used by Wiesenfeld\cite{H2-H2O-Wiesenfeld-2022} which consists in replacing  $ \varGamma_{\nu'_{1} \nu'_{2} \nu'_{3}} ^{j'_{1}\tau'}  (R_{1},R_{2},\alpha)$ by $ \varGamma_{\nu'_{1} \nu'_{2} \nu'_{3}} ^{0 1}  (R_{1},R_{2},\alpha)$ and $ \varGamma_{\nu_{1} \nu_{2} \nu_{3}} ^{j_{1}\tau}  (R_{1},R_{2},\alpha)$  by $ \varGamma_{\nu_{1} \nu_{2} \nu_{3}} ^{0 1}  (R_{1},R_{2},\alpha)$  in equation( \ref{eqn:pot}).

We represented in the lower panel of Figure \ref{fig:ratio} the ratio of the present Close Coupling (denoted $ \sigma $ )and approximate state to state cross section  (denoted $ \sigma^{DP} $ for direct product) as a function of collision energy for the collisions of  \textit{p-}\ce{H2} with 
\ce{H2O}$ (\nu_{2}=1,3_{31}) $. As can be seen on this figure the ratio of the Close Coupling and approximate state to state cross sections is ranging from 0.4  to 2.2. The coupling is seen not surprisingly to be the strongest for collision energies associated with the region of the well and the long range part of the PES. The strongest peaks which can be seen on this figure are associated with transitions towards the fundamental bending level: \textit{p-}\ce{H2}(j$ _{2} $=4)+ \ce{H2O}$ (\nu_{2}=0,5_{24}) $; \textit{p-}\ce{H2}(j$ _{2} $=2)+ \ce{H2O}$ (\nu_{2}=0,5_{42}) $; \textit{p-}\ce{H2}(j$ _{2} $=0)+ \ce{H2O}$ (\nu_{2}=0,5_{24}) $ and with $ \Delta j_{2}=2 $. Conversely the strongest bump at very low collision energy is associated with a pure rotational transition inside the excited \ce{H2O} bending level towards  \textit{p-}\ce{H2}(j$ _{2} $=0)+ \ce{H2O}$ (\nu_{2}=1,3_{22}) $. \\

If we now  examine  the ratio of the global bending relaxation cross section of \ce{H2O}$ (\nu_{2}=1,3_{31}) $ by collision with \textit{p-}\ce{H2}(j=0) represented in the upper panel of Figure \ref{fig:ratio} as a function of collision energy  we find that the global error is divided by a factor two as the ratio is ranging from 0.7 to 1.4. The curve exhibits a succession of peaks located in the collision energy interval associated with the PES well, demonstrating clearly the resonant character of the coupling between bending and rotation.

\section{Conclusion}\label{Concl}

We presented the first calculations of the bending relaxation of
\ce{H2O} by collisions with the two spin modifications of \ce{H2} when
including exactly the coupling between \ce{H2O} bending and
rotation. The approach presented also considers the \ce{H2O} stretching
modes while only the two lowest bending levels
were included in the present study. Our approach allowed
us to obtain several important results. First, for collision energies
associated with the region of the intermolecular potential well,
roughly between 10 and 300 K,  bending relaxation is
controlled by a near-resonant energy transfer mechanism. Consequently,
bending relaxation is larger towards the most excited final open
rotational level $j_2= 4$ and decreases monotonously when $j_2$
decreases for collisions with \textit{p-}\ce{H2}. In the case of
collisions with \textit{o-}\ce{H2} it is the $j_2= 3$ final channel
which is dominating as the $j_2= 5$ level is closed for several
initial levels of the \textit{o-}\ce{H2} + \ce{H2O}$ (\nu_2=1,j
k_{A}k_{C}) $ system. The Boltzmann weights make furthermore
the contribution of the collisions with \ce{H2}(j$ _{2}=1,2 $) 
negligible at low and moderate temperature. A very good approximation (10\% error) of the global bending relaxation rate can 
be obtained by considering the process to be
controlled by the collisions with \textit{p-} \ce{H2} for the whole
interval of temperature of the present study ($0.5-500~K$). 
Collisions with \textit{o-}\ce{H2} would do not need in this case to be computed
thus dividing computer time by more than a factor of two. \\ 
Our computed value of the global bending
relaxation rate coefficient at 295 K is in excellent agreement with the experimental
value. Future (ideally state-resolved) measurements over an extended
range of temperatures will be useful to further test the predictions of the
present calculations.We eventually investigated the decoupling between bending and rotation and found that the use
of such an approximation leads to error ranging from 50\% to 100\%  respectively for the bending relaxation and state to state cross sections.

\end{doublespacing}

\section*{Acknowledgments}

This work was supported by the french Agence Nationale de la Recherche
(ANR-Waterstars), Contract No. ANR-20-CE31-0011 and by the ECOS-SUD
project C22E02. Most of the computer time for this study was provided
by the \emph{M\'esocentre de Calcul Intensif Aquitaine} computing
facilities of \emph{Universit\'e de Bordeaux and Universit\'e de Pau
	et des Pays de l'Adour}. Part of the computations were performed using
the GRICAD infrastructure (https://gricad.univ-grenoble-alpes.fr),
which is supported by Grenoble research communities.

\section*{Conflicts of interest}
There is no conflict of interest to report.
\renewcommand\refname{References}
\begin{footnotesize}
\bibliographystyle{unsrt.bst} 
\textnormal{\bibliography{main}}
\end{footnotesize}
\newpage


\end{document}